# A multifunctional metasurface: from extraordinary optical transmission to extraordinary optical diffraction in a single structure

Zi-Lan Deng[1], Xiangping Li[1,*], and Guo Ping Wang[2,*]

[1]Guangdong Provincial Key Laboratory of Optical Fiber Sensing and Communications, Institute of Photonics Technology, Jinan University, Guangzhou 510632, China
[2]College of Electronic Science and Technology, Shenzhen University, Shenzhen 518060, China.

*Email: xiangpingli@jnu.edu.cn; gpwang@szu.edu.cn

**Abstract**

We show that, a metasurface composed of subwavelength metallic slit array embedded in an asymmetric environment can exhibit either extraordinary optical transmission (EOT) or extraordinary optical diffraction (EOD). By employing an analytical model expansion method and the diffraction order chart in k-vector space, we found that the resonance decaying pathway of the local slit cavity mode can be tuned to either $0^{th}$ or $-1^{st}$ diffraction order by changing the parallel wavevector, which gives rise to enhanced $0^{th}$ transmission (EOT) of the structure for small incident angles, and enhanced $-1^{st}$ diffraction (EOD) for large incident angles. Based on this appealing feature, a multifunctional metasurface that can switch its functionality between transmission filter, mirror and off-axis lens is demonstrated. Our findings provide a convenient way to construct multifunctional integrated optical devices on a single planar device.

Integrating various functionalities on a single miniaturized photonic device can largely simplify the overall optical system design, and therefore is a continuing trend in the nanophotonic area. There are many ways to realize multifunctional devices, such as the so called "Janus" transformation optics device [1], acting as either a lens or a beam-shifter; the parity-time-symmetric multilayers that can simultaneously support extraordinary transmission and reflection [2]; a single parity-time-symmetric cavity that can support both lasing and anti-lasing [3] and so on. Those demonstrations require large coherent interaction lengths during light's propagation in such devices.

Recently, exploring various wavefront-shaping functionalities in a single ultrathin layer, i.e., metasurface, has been widely studied in nanophotonic field due to the easy fabrication of a flat monolayer instead of a three-dimensional bulky structure [4-6]. Resorting to the abrupt phase change of each resonant element instead of accumulated interaction with large coherent length between light and structure, ultrathin metasurface can provide various applications, such as beam deflection [7-11], focusing [12-16], complex beam shaping [17-21], holographic display [22-26], and multiplexed hologram recording [27-36]. Generally, the previous metasurface can only work on either transmission mode [7] or reflection mode [9,25] depending on deterministic decaying pathway of its resonance. For transmission metasurfaces, only the transmission decaying channel is used for wavefront manipulation, while the reflection decaying channel is abandoned. For reflection metasurfaces, a metallic mirror is usually placed below the metasurface nanostructure to block all transmission channels, and only the reflection channel is tailored on demand. In this paper, we present a metallic slit metasurface that harnesses both the transmission and reflection decaying channels of different diffraction orders, and demonstrate different functionalities for different resonance decaying channels.

It is well known that, for a metallic slit array with subwavelength period ($p<\lambda$), the localized resonance mode in the slit area leads to the enhanced transmission of incident light, which is known as extraordinary optical transmission (EOT) [37,38]. To obtain enhanced diffraction of a grating, the usual way is to mimic a blazed grating by making a gradient phase profile with discrete set of different nano-resonators in a large supercell [7,39,40]. Recently, it is also shown that, rather than a blazed grating approach, in a metallic binary slit array with period comparable with working wavelength ($p\sim\lambda$) that is blocked by a metallic background, the localized slit resonance mode can lead to the enhanced diffraction of light in the -1st diffraction order to near-unity, which is refer to as extraordinary optical diffraction (EOD) [20]. Now, we consider a metallic subwavelength slit array embedded in an asymmetric background medium without a metallic ground plate, and demonstrate that both the EOT and EOD can happen in a single structure. According to the diffraction order chart in k-vector space, the resonance decaying pathways of the slit cavity mode can be tuned to either 0th transmission or -1st reflection channels by changing the parallel wavevector of incident light. As a result, the metasurface exhibits resonantly enhanced transmission for small incident angles, acting as a transmission filter; exhibits total internal reflection (TIR) for moderate incident angles, acting as a refection mirror; and exhibits EOD for large incident angles, which can be applied for arbitrary wavefront shaping, like deflecting, focusing, beam-shaping and hologram displaying. Our proposed metasurface can control both the transmission and reflection channels of impinging light by incident angles, providing a convenient and efficient way to realize multifunctional integrated optical devices on a single surface.

Figure 1(a) illustrates the schematic of the multifunctional metasurface, which is a metallic subwavelength slit array (with slit width $w$, slit height $h$, and period $p$) embedded in an asymmetric environment. The refractive index of the medium in the incident side is larger than that of the medium in the transmission side ($n_1>n_3$) to make sure that there are intersection areas that are not only higher than $0^{th}$ reflection Woods anomaly (RWA) but also lower than $0^{th}$ transmission Woods anomaly (TWA) [green and red patches in Fig. 1(b)]. The index in the slit area $n_2$ is chosen properly so that the resonance wavelength of the slit cavity mode can cross all the three kinds of areas with different diffraction order combinations that is marked as blue, green and red patches in Fig. 1(b). For small, moderate and large incident angles, the beam will undergo EOT, TIR and EOD, respectively, as the decaying pathways of the slit cavity mode are composed by different diffraction order combinations for different angles. Assume a plane wave with wavenumber $k_0$ and parallel wavevector $k_{x0}$ illuminating the structure from the superstrate, the possible propagating diffraction orders for both reflection ($r_0$, $r_{\pm 1}$, …) and transmission ($t_0$, $t_{\pm 1}$, ...) can be found in the diffraction order chart as shown in Fig. 1(b). The diffraction order chart is constructed by RWA (solid) and TWA (dashed) lines, which are governed by Equations $k_{x0}+m2\pi/p=\pm n_1 k_0$ and $k_{x0}+m2\pi/p=\pm n_3 k_0$ ($m$ is an integer), respectively. Above the $m^{th}$ ($m$=0, ±1, ±2, …) RWA/TWA, the propagating reflection/transmission of the $m^{th}$ diffraction order ($r_m$/$t_m$) begins to appear, while below the $0^{th}$ RWA, which is also known as the light line, there are only evanescent waves without any radiation. For simplicity, we only consider the lowest areas of the diffraction order chart, where only a few diffraction orders (0 and ±1) exist. In the lowest sidelobe areas (green patches), there is only $r_0$ channel, where TIR happens; in the lowest central area (blue patch), there are only two propagating channels of $r_0$ and $t_0$, which may give rise to EOT; and in the higher sidelobe areas (red patches), there are only two propagating channels of $r_0$ and $r_{-1}$/$r_1$ allowing for the possibility of EOD. As the diffraction order chart is symmetric with respect to the $k_0$ axis, we only consider the $k_{x0}>0$ case in the following discussions.

To obtain the transmission and reflection spectra of different diffraction orders, we first employ the analytical model expansion theory by considering the metal as perfect electric conductor (PEC) [20,41]. When a plane wave with parallel wavevector $k_{x0}=n_1 k_0 \sin\theta$ and transverse magnetic (TM) polarization illuminates on the metallic slit array, by matching the boundary conditions which require that the tangential components of both electric field ($E_x$) and magnetic field ($H_y$) are continuous at interfaces, we can analytically obtain the $0^{th}$ reflectance $R_0$, $0^{th}$ transmittance $T_0$, and -$1^{st}$ reflectance $R_{-1}$, as follows (see Appendix A of supporting material),

$$R_0 = \left| \frac{\varepsilon_1}{\varepsilon_2} \frac{n_2 k_0}{k_{z0}^{(1)}} \frac{w}{p} \sin c\left(\frac{k_{x0}w}{2}\right)(a_0^+ + a_0^-) - 1 \right|^2, \tag{1a}$$

$$T_0 = \left| e^{-ik_{z0}^{(3)}h} \frac{\varepsilon_3}{\varepsilon_2} \frac{n_2 k_0}{k_{z0}^{(3)}} \frac{w}{p} \sin c\left(\frac{k_{x0}w}{2}\right)(a_0^+ e^{-in_2 k_0 h} + a_0^- e^{in_2 k_0 h}) \right|^2, \tag{1b}$$

$$R_{-1} = \left| \frac{\varepsilon_1}{\varepsilon_2} \frac{n_2 k_0}{k_{z,-1}^{(1)}} \frac{w}{p} \operatorname{sinc}\left(\frac{k_{x,-1} w}{2}\right) \left(a_0^+ + a_0^-\right) \right|^2, \qquad (1c)$$

Where, $a_0^+$ and $a_0^-$ are forward and backward complex amplitude coefficients of the cavity mode in the subwavelength slit area, which can be obtained by solving linear equations

$$\begin{pmatrix} a_0^+ \\ a_0^- \end{pmatrix} \begin{pmatrix} m_{11} & m_{12} \\ m_{21} & m_{22} \end{pmatrix} = \begin{pmatrix} 2\operatorname{sinc}(k_{x0} w/2) \\ 0 \end{pmatrix}, \qquad (2)$$

with

$$m_{11} = \frac{\varepsilon_1}{\varepsilon_2} \frac{w}{p} \sum_{m=-\infty}^{\infty} \frac{n_2 k_0}{k_{zm}^{(1)}} \operatorname{sinc}^2\left(\frac{k_{xm} w}{2}\right) + 1, \qquad (3a)$$

$$m_{12} = \frac{\varepsilon_1}{\varepsilon_2} \frac{w}{p} \sum_{m=-\infty}^{\infty} \frac{n_2 k_0}{k_{zm}^{(1)}} \operatorname{sinc}^2\left(\frac{k_{xm} w}{2}\right) - 1, \qquad (3b)$$

$$m_{21} = e^{in_2 k_0 h} \left[ \frac{\varepsilon_3}{\varepsilon_2} \frac{w}{p} \sum_{m=-\infty}^{\infty} \frac{n_2 k_0}{k_{zm}^{(3)}} \operatorname{sinc}^2\left(\frac{k_{xm} w}{2}\right) - 1 \right], \qquad (3c)$$

$$m_{22} = e^{-in_2 k_0 h} \left[ \frac{\varepsilon_3}{\varepsilon_2} \frac{w}{p} \sum_{n=-\infty}^{\infty} \frac{n_2 k_0}{k_{zm}^{(3)}} \operatorname{sinc}^2\left(\frac{k_{xm} w}{2}\right) + 1 \right]. \qquad (3d)$$

where, ($k_{x0} = k_0 \sin\theta$, $k_{z0}^{(1)} = \sqrt{n_1^2 k_0^2 - k_{x0}^2}$) is the wavevector of incident plane wave, and ($k_{xm} = k_{x0} + 2m\pi/p$, $k_{zm}^{(j)} = \sqrt{n_j^2 k_0^2 - k_{xm}^2}$) ($j$=1, 3) is the wavevector of the $m^{\text{th}}$ diffraction order in superstrate and substrate, respectively.

Figures 2(a)-2(c) show the analytically obtained $R_0$, $T_0$ and $R_{-1}$ for varying wavelengths and incident angles, by Eqs. (1a-1c), respectively. The parameters of the metallic slit array are $p$=1um, $w$=0.1um, $h$=1um; refractive indices of the upper layer, the slit area and lower layer are $n_1$=2, $n_2$=1.4, and $n_3$=1, which can be realized by $Si_3N_4$, $SiO_2$, and air, respectively. In Fig. 2(a), there are three abrupt color changing contours, which correspond to the $0^{\text{th}}$ TWA (critical angle for TIR), -$1^{\text{st}}$ RWA and -$1^{\text{st}}$

TWA in Fig. 1(b), respectively. The subarea ① (upper-left), ② (upper-middle), and ③ (upper right) in Fig. 2(a) separated by those WA lines corresponds to the blue patch, green patch and red patch in Fig. 1(b), respectively. Thus, there are only $r_0$ and $t_0$ channels in ①, only $r_0$ in ②; and only $r_0$, $r_{-1}$ channels in ③, respectively. We see that, $R_0$ exhibit dips in subareas ① and ③, at the resonant wavelength of about 3.3 μm, which corresponds to the fundamental cavity mode in the metallic silts. For different incident angles, the resonant wavelength stays the same, because the slit cavity mode that leads to the resonant dip of $R_0$ is a localized resonant mode, whose excitation is independent of the incident angles. Correspondingly, $T_0$ exhibits a resonant peak approaching unity at the same position of the $R_0$ dip in subarea ① ($\theta_0 < asin(n_3/n_1) = 30°$) as shown in Fig. 2(b). Note that, the subarea ① is the widely studied zeroth-order-diffraction zone in the metamaterial community, many phenomena like EOT [37], Generalized Brewster effect [42-44], perfect absorption [45] are found in this zone, while the influence of the resonance behavior on other areas other than subarea ① receives scarce attentions. When $\theta_0 > 30°$, $T_0$ is always zero because the incident angle is larger than the critical angle for TIR. Figure 2(c) shows that, $R_{-1}$ also exhibits a peak at the same position of the $R_0$ dips in subarea ③. Therefore, the slit cavity mode indeed can enhance the diffraction efficiency of the -1$^{st}$ order to near unity, as long as the propagating channels are restricted to only $r_0$ and $r_{-1}$. We note that, between the EOT and EOD, there is a transitional area (subarea ②), where $R_0$ is constantly unity [Fig. 2(a)]. This is due to the fact that, $r_0$ is the only propagating channel in this subarea, the resonant behavior of the slit cavity can only lead to a phase change of the reflection coefficient, while the reflection amplitude always stays unity. However, when the absorption loss channel is considered in the practical case, there may be resonant absorption phenomenon that can also lead to the $R_0$ dip in subarea ②.

According to the theory we discussed above, now let us consider the practical case of real metallic structures. Silver is chosen as the metal material, with permittivity obtained by fitting the experimental data [46] to the Drude model. Figure 3 shows the $R_0$, $T_0$, and $R_{-1}$ spectra by finite element method (FEM), which is implemented by a commercial package COMSOL. At incident angles of 0º [Fig. 3(a)] and 15º [Fig. 3(b)] that belong to the subarea ①, $R_0$ (green) exhibits resonant dips below 0.1 while $T_0$ (blue) exhibits peaks exceeding 0.9 at the wavelength of about 3.3um, which is the typical EOT phenomenon. At incident angles of 45º [Fig. 3(c)] and 60º [Fig. 3(d)] that belong to the subarea ③, $R_0$ (green) also exhibits resonant dips at the wavelength around 3.2um, while $R_{-1}$ (red) exhibits resonant peaks exceeding 0.9 at the corresponding wavelength, which is the EOD phenomenon. We note that, there are sharp peaks for $R_{-1}$ and sharp dips for $R_0$ with Fano lineshape at the short wavelength side in Figs. 3(b)-3(d), which is due to the Wood's anomalies near those wavelengths. Compared with the model expansion theory results (dashed lines in Fig. 3), the peak values of $T_0$ and $R_{-1}$ of the FEM simulation results are slightly lower, and the resonant wavelength shift a little to shorter wavelengths, because the real metal inevitably has absorption loss and finite negative permittivity, which give rise to the overall power loss and the decreasing of effective wavelength of the slit mode. Nevertheless, the FEM simulation results for real metal [solid lines in Fig. 3] are generally consistent with the theoretical results [dashed lines in Fig. 3]. The appearance of the EOT and EOD relies on the different resonant decaying pathways ($0^{th}$ or $-1^{st}$ diffraction order channels) of the local slit cavity mode, therefore we can tailor the working wavelength and bandwidth of EOT and EOD through the slit parameters. In general, the resonant wavelength is linearly dependent on the slit height $h$, and the bandwidth is positively correlated to the slit width w (see Appendix B of the supporting material). As a result, our proposed structure provides a flexible way to tailor the working condition of metasurface on demand.

To clearly illustrate the evolution from EOT to EOD, we fix the working wavelength at 3.3um, and plot the $R_0$, $T_0$ and $R_{-1}$ as a function of the incident angle as shown in Fig. 4(a). It indicates that, as we gradually increase the incident angle from 0º to 90º, the beam first undergoes EOT ($T_0>0.8$; $R_0$, $R_{-1}<0.1$). And after the critical angle 30º, the beam experiences TIR, where $R_0$ is the only nonzero component. To further increase the incident angle to cross the $-1^{st}$ RWA line ($\theta>asin[\lambda/(n_1 p)-1]=40º$), the EOD begins to happen, where $R_{-1}$ becomes the largest component, while $R_0$ and $T_0$ are effectively suppressed. Figures 4(b)-4(f) show the field pattern ($|H_z|$) for different incident angles. We can clearly see that, for small incident angles (0º, 15º), the beam passes through the metallic slit metasurface [Figs. 4(b) and 4(c)], even if the slit width is in deep subwavelength range (w=0.1um<<$\lambda$=3.3um), and the relation between the transmission angle $\theta_t$ and incidence angle $\theta$ obeys the Snell law ($n_1 sin\theta = n_3 sin\theta_t$). For incident angle 35º in subarea ②, the incident beam undergo the ordinary TIR, with the reflection angle equal to the incident angle [Fig. 4(d)]. For large incident angles (45º, 55.6º) in subarea ③, the incident beam undergo the negative reflection, with the reflected beam on the same side of the incident beam with respect to the normal [Figs. 4(e) and 4(f)], which is the typical behavior of EOD. We note that, the restro-reflection happens when the incident angle is 55.6º (Littrow mount angle), that is, the beam is reflected back along its original path, even for oblique incidence.

Based on EOD, we can design arbitrary wavefront-shaping metasurfaces with the binary computer-generated hologram approach [47]. As an example, we design a multi-functional metasurface that can behave as an off-axis lens in the EOD domain by modulating the metallic slit array with a quadratic phase profile $\phi(x,y) = k_0 x^2 / 2f$. The structure of this metasurface is shown in Fig. 5(a), exhibiting a linear gradient period profile. For small incident angles [Figs. 5(b) and 5(c)], the beam still can pass through the non-periodic subwavelength slit array, exhibiting the EOT phenomenon. For the incident angle of 35º [Fig. 5(d)], the metasurface acts as an ordinary mirror, making the incident beam undergo the TIR. For large incident angles [Figs. 5(e) and 5(f)], the metasurface becomes an off-axis lens, which can focus the incident beam into a single line in the -1$^{st}$ diffraction direction, while the specular reflection in the right side is near-totally suppressed. Therefore, for different incident angles, the metasurface indeed can act as an optical device with totally different functionalities. This angle-dependent mutli-functional device may be potentially useful in panoramic viewing systems.

In summary, we found that a metallic subwavelength slit array embedded in asymmetric environment can support EOT, TIR and EOD simultaneously. This is achieved by angularly tailoring resonance decaying pathways of the metallic slit cavity mode to different diffraction orders. For small incident angles, the resonance decaying pathway goes to the $0^{th}$ transmission channel, giving rise to enhanced transmission to near unity, while for large incident angles, resonance decaying pathway switches to $-1^{st}$ reflection channel, leading to enhanced diffraction to near unity. Between the EOT and EOD area, there is also a transitional area with moderate incident angles, where ordinary specular reflection happens. Based on the above theory, a multifunctional metasurface that can behave as a transmission filter, mirror and an off-axis lens simultaneously is demonstrated. Our proposed metasurface can be used to realize multifunctional integrated optical devices on a single planar device.

This work is supported by National Natural Science Foundation of China (NSFC) (Grant 11604217, 11574218, 61522504, 61420106014).

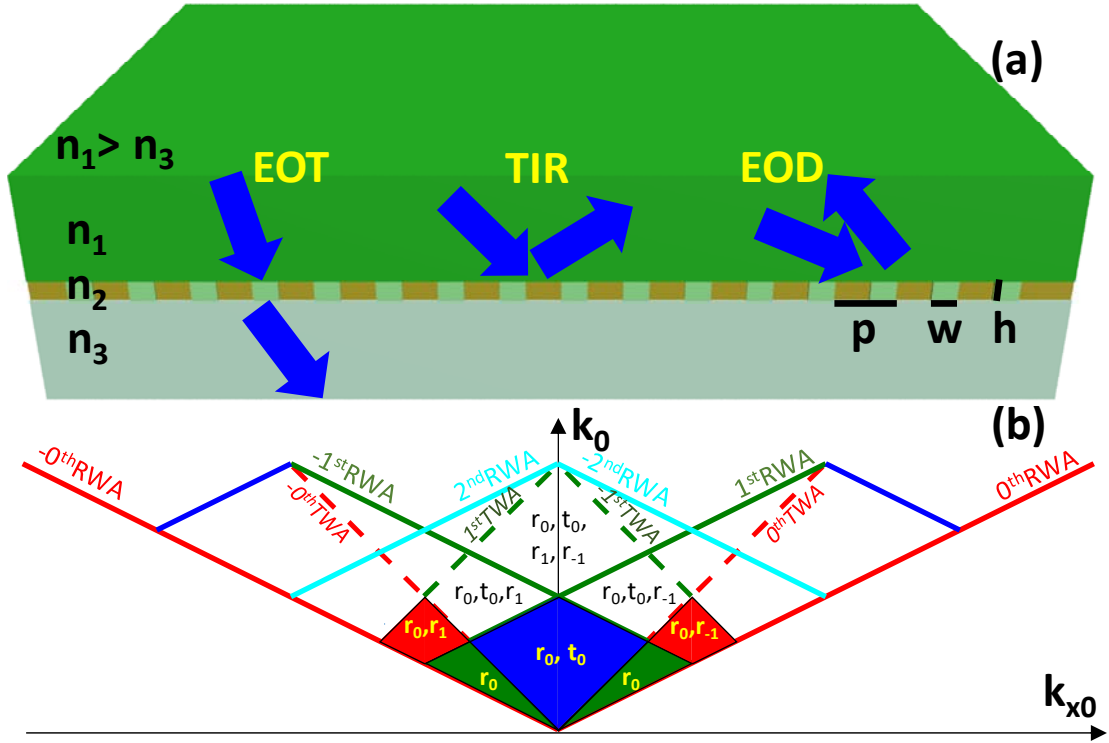

FIG. 1. (color online) (a) Schematic of the metasurface structure that can exhibit EOT, TIR or EOD, depending on the incident angle. The index of the upper layer, slit area, and lower layer is $n_1 = \sqrt{\varepsilon_1 \mu_1} = 2$, $n_2 = \sqrt{\varepsilon_2 \mu_2} = 1.4$, and $n_3 = \sqrt{\varepsilon_3 \mu_3} = 1$, respectively. (b) The diffraction order chart in k-vector space of the structure in (a). Solid lines represent the RWAs with orders ±0, ±1, ±2, while the dashed lines represent TWAs with orders ±0, ±1.

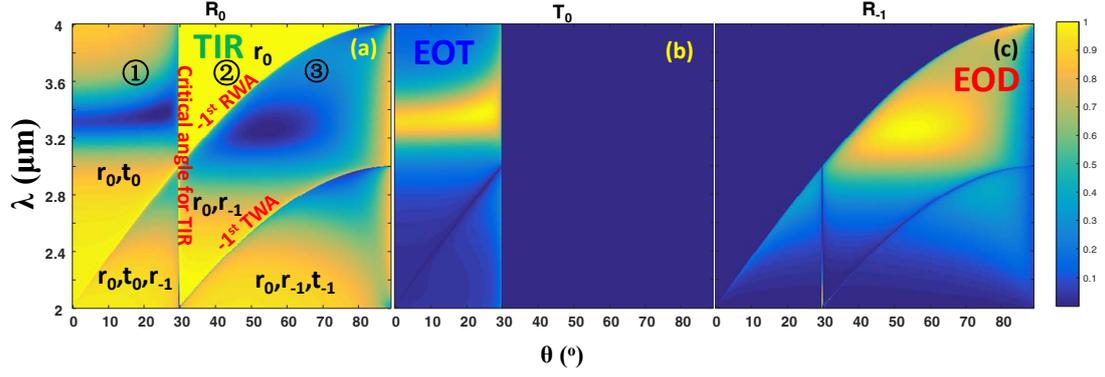

FIG. 2 (color online). Analytical results of the (a) $0^{th}$ reflectance $R_0$, (b) $0^{th}$ transmittance $T_0$ and (c) $-1^{st}$ reflectance $R_{-1}$, respectively, of the metasurface for varying incident angle $\theta$ and wavelength $\lambda$. The three abrupt color changing contours in (a) automatically show the critical angles for TIR, the $-1^{st}$ RWA, and the $-1^{st}$ TWA, respectively, as marked in the figure, and they split the whole phase map into different subareas that contain combinations of different propagating channels (also denoted in the figure). In subarea ①, there are $r_0$ and $t_0$ channels, $R_0$ exhibits a dip in (a) while $T_0$ exhibits a peak in (b) at the resonance condition of the slit cavity mode. In subarea ②, there is only $r_0$ channel, $R_0$ is unity in the whole subarea. In subarea ③, there are $r_0$ and $r_{-1}$ channels, $R_0$ exhibits a dip in (a) while $R_{-1}$ exhibits a peak in (c) at the resonance condition of the slit cavity mode.

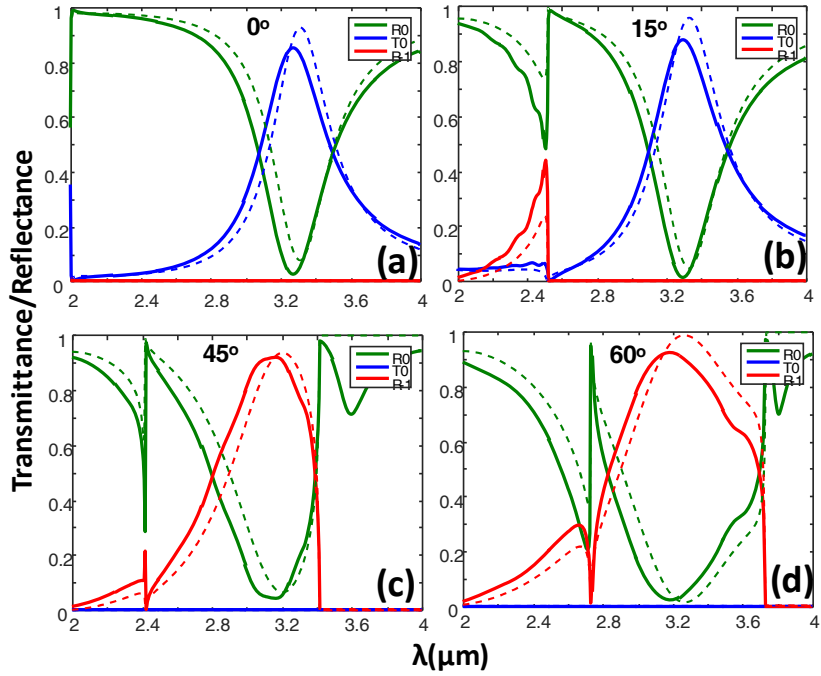

FIG. 3 (color online) Reflection ($R_0$, $R_{-1}$) and transmission ($T_0$) spectra of the metasurface composed of real metal (silver) with parameters p=1μm, h=0.8μm, w=0.1μm, for different incident angles (a) θ=0º, (b) θ=15º, (c) θ=45º, (d) θ=60º, respectively. Dashed curves show the corresponding analytical results of PEC metasurface with p=1μm, h=1μm, w=0.1μm for comparison.

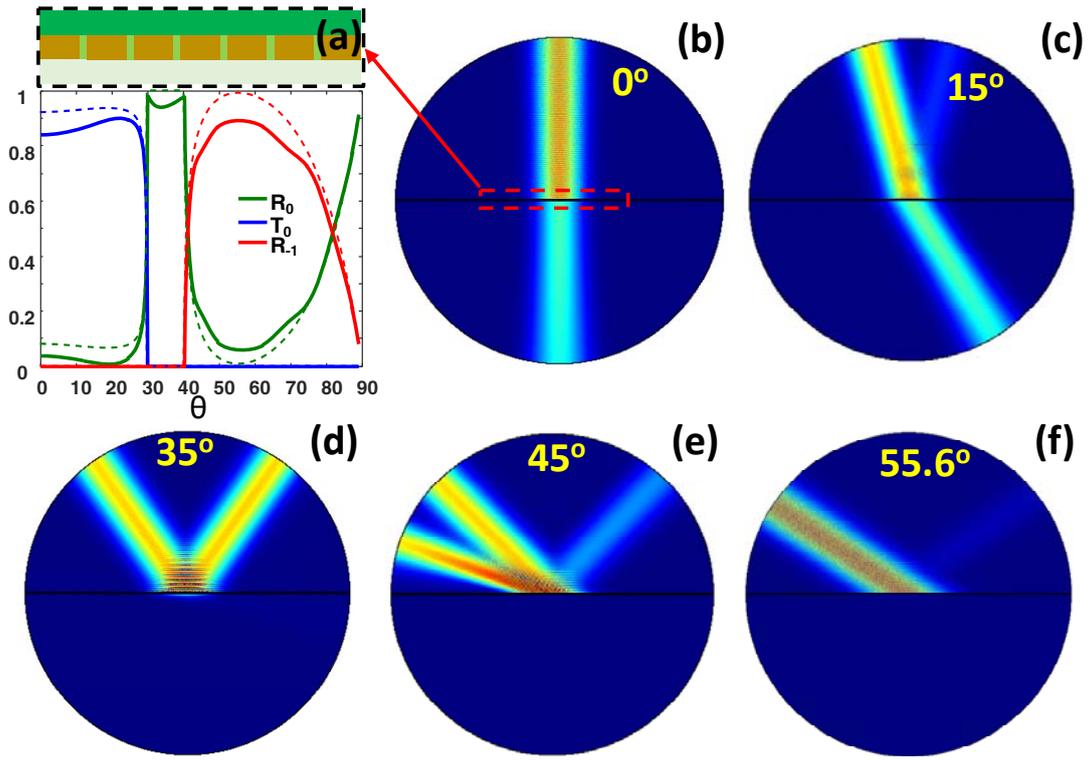

FIG. 4 (color online) (a) Reflection ($R_0$, $R_{-1}$) and transmission ($T_0$) as a function of the incident angle $\theta$ at the working wavelength $\lambda=3.3\mu m$ for the periodic slit array of which the zoom-in structure is shown in the upper panel. Dashed curves show the corresponding analytical results of PEC metasurface for comparison. (b-f) Field patterns ($|H_z|$) of a Gaussian beam with $\lambda=3.3\mu m$ illuminating the metasurface (the area marked by the red dashed frame) with different incident angles: (b) 0º, (c) 15º, (d) 35º, (e) 45º, and (f) 55.6º, respectively.

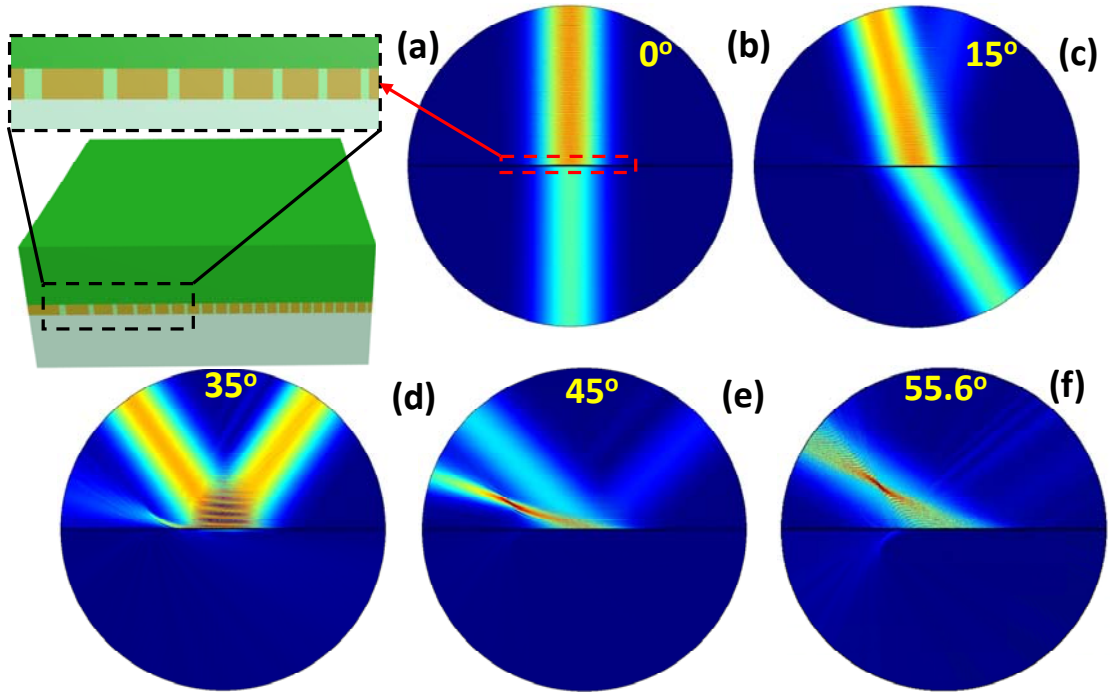

FIG. 5 (color online) (a) Schematic of the modulated metasurface with a gradient grating profile with a zoomed-in structure on the upper panel. (b-f) Field patterns ($|H_z|$) of a Gaussian beam with λ=3.3μm illuminating the modulated metasurface (the area marked by the red dashed frame) with different incident angles: (b) 0º, (c) 15º, (d) 35º, (e) 45º, and (f) 55.6º, respectively. In (b, c), the incident beam also undergoes EOT. In (d), the incident beam still undergoes TIR. While in (e) and (f), the incident beam undergoes the focusing capability, with near-total suppression of the $0^{th}$ diffraction order.